# AxLLM: Accelerator Architecture for Large Language Models with Computation Reuse Capability

Soroush Ahadi[1], Mehdi Modarressi[1], Masoud Daneshtalab[2,3]

soroush.ahadi@ut.ac.ir, modarressi@ut.ac.ir, masoud.daneshtalab@mdu.se

[1]School of Electrical and Computer Engineering, University of Tehran, Tehran, Iran
[2]Intelligent Future Technologies (IFT), Mälardalen University, Västerås, Sweden
[3]Department of Computer Systems, Tallinn University of Technology (TalTech), Tallinn, Estonia

**Abstract**. Large Language Models (LLMs) demand massive computational power and memory resources, posing significant challenges for efficient deployment. While quantization has been widely explored to reduce model size and computation, this paper demonstrates an additional benefit: quantization increases parameter locality, creating opportunities for computation reuse. Building on this insight, we propose AxLLM, a hardware accelerator architecture designed for quantized LLMs. AxLLM introduces a novel redundancy elimination technique that caches and reuses multiplication results for repeated weight values, substantially reducing redundant operations. The architecture features dual "multiply" and "reuse" pipelines, efficiently supporting both base models and LoRA fine-tuned models without altering parameters, retraining, or requiring offline preprocessing. Experimental results show that AxLLM achieves up to 90% reduction in computations, delivering 28% lower energy consumption and a 1.7× speedup over baseline execution. These results highlight AxLLM as a scalable and efficient solution for accelerating LLMs on specialized hardware.

## I. Introduction

The Large Language Models (LLMs) are neural networks, in the form of Transformer architecture, trained on vast amounts of textual data for natural language processing tasks, such as understand and generate human-like language. LLMs are heavy in terms of computation and memory requirements. For example, GPT-3 and Google's PaLM have 175 billion and 540 billion parameters, respectively [1], demanding extensive computational power and hundreds of gigabytes of memory and disk space. These demands underscore the challenges of running LLMs and, much like other deep learning (DL) models [2], highlight the necessity of specialized hardware design for transformer models [1][3-4].

Quantization is base of most existing hardware accelerators for transformers: many previous studies have shown that the parameters and data of many transformers can be quantized to low bit widths, such as 8-bit or even lower, to reduce memory usage and computation load without significantly reducing their accuracy [1,3-5].

In this paper, we observe that, in addition to reducing computational load and model size, quantization provides a secondary benefit by increasing the locality of values in model parameters. It increases the frequency of repetitive parameters, which in turn, leads to a higher rate of repetitive computation. To take advantage of this opportunity, we present AxLLM, an accelerator architecture designed for quantized LLMs. By leveraging computation reuse to eliminate redundancy, AxLLM significantly reduces the computational load of LLMs and effectively boosts their execution speed.

In our previous work, we demonstrated that eliminating redundant computations in CNNs [6] and MLPs [7] significantly improves power efficiency and reduces latency. Building on this foundation, this paper introduces a novel redundancy elimination technique tailored specifically for transformers, along with the AxLLM accelerator architecture to implement it.

AxLLM exploits the vector-to-matrix dataflow in transfromers, where each input vector element multiplies all weight matrix cells in its corresponding row. In large LLM matrices with low-bit quantization (e.g., 8-bit), The large number of cells in each row can take only 256 distinct values. AxLLM employs an input-stationary order, where input cells are fetched and processed sequentially. For each input, AxLLM caches the multiplication results of unique values upon first occurrence and reuses them for subsequent appearances. So, the AxLLM has two parallel "multiply" and "reuse" pipelines: A lightweight logic block determines whether this is the first occurrence of a unique weight and routes the operands to the appropriate pipeline.

AxLLM also works for Low-Rank Adaptation (LoRA) fine-tuned LLM models. Fine-tuning adapts a pre-trained LLM to a specific task or domain by training it on tailored datasets, enhancing its performance in that area. LORA is a fine-tuning method for LLMs that adds minimal memory overhead: it adapts pre-trained models to specific tasks by injecting small, trainable low-rank matrices, called adaptor matrices, into the base model layers [8]. AxLLM efficiently reuses computations from the base weight matrix to also process LORA adapter elements with matching values, significantly reducing the additional computational overhead introduced by adaptation.

A key advantage of AxLLM over existing LLM accelerators is that it preserves the original parameters and hyperparameters of the language model. Unlike many approaches that rely on fine-tuning or retraining to mitigate accuracy degradation, AxLLM is a post-training mechanism, effectively avoiding the costly and time-consuming retraining processes entirely. Furthermore, AxLLM, unlike some previous work [9][10], requires no offline phase to process the weights or precompute results, so it has zero setup time.

Experimental results show up to 90% reduction in LLM calculations, which translates to up to 28% energy reduction and 1.7x boost in speed.

## II. RELATED WORK

Prior LLM acceleration work can broadly be classified into four categories: (1) data-movement–aware methods (memory and near-/in-memory) that reduce or hide bandwidth/capacity pressure via KV-cache and activation quantization/compression, offloading across tiers, layout/tiling optimizations, and moving parts of calculations closer to storage; (2) approximate/decoding-aware methods that cut per-token work with speculative decoding, early exit, and draft-and-verify pipelines; (3) low-bit and multiplier-reduction compute that replaces multiplications with cheaper primitives (quantization, shift–add, lookup tables, bit-serial paths); and (4) algorithm–architecture co-design introducing new instructions, tensor-core styles, and accelerator microarchitectures tailored to attention and matrix multiplication.

AxLLM can be classified as a multiplier-reduction/eliminations method. It is orthogonal and can be used along with many memory-aware and approximate/decoding methods.

There are several outstanding multiplier-reduction methods in the literature. Ax an example, ShiftAddLLM exemplifies the multiplier-reduction class with a post-training reparameterization that decomposes weights into scaled binary bases so that matrix multiplications become shift-and-add operations [9]. A light calibration aligns outputs without fine-tuning, enabling multiplier-free execution while retaining competitive accuracy and efficiency.

The LUT Tensor Cores work is an algorithm–architecture co-design for mixed-precision GEMM that organizes computation around compact lookup tables. It introduces a tensor-core–style LUT unit that maximizes table reuse via elongated tiling, employs a bit-serial datapath to flexibly mix precisions, and provides ISA/compiler support to minimize table generation and movement—improving compute density and energy efficiency over prior LUT approaches [10].

Oaken method [3] targets the data-movement axis by addressing KV-cache bottlenecks with an online–offline hybrid quantization scheme: outlier thresholds are set offline, then used online to pick scales cheaply at runtime. Coupled with dedicated quantization engines and memory management, Oaken reports up to 1.58× throughput at batch-256 with only ~0.54% average accuracy loss versus state-of-the-art KV-quantization [3].

Early-exit hybrids such as EESD [11] fuse early exit with speculative decoding: they reuse the first $N$ layers of the same LLM as a drafter (sharing the KV cache, so no second model is needed), add a thin layer and head trained via self-distillation, adapt draft length per prompt (e.g., with Thompson sampling), and keep outputs exact via verification.

AxLLM is post-training (no retraining or fine-tuning), achieves no accuracy loss, and accelerates inference through explicit computation reuse—exploiting repeated sub-computations rather than approximating or changing numeric formats. It imposes very low overhead in storage and runtime control, avoiding large auxiliary tables and complex online procedures, and preserves exact arithmetic semantics while complementing the data-movement–aware and approximate/decoding-aware lines of work.

The benefits of computation reuse have already been explored in other AI models, including in some of our prior work [6][7][12], but to the best of our knowledge, AxLLM is the first to apply computation reuse to Transformers.

There are some recent works on reuse in LLMs at a coarser granularity: as an example, AttMemo [13] caches self-attention results and reuses them when inputs are sufficiently similar, controlling reuse via an embedding-based similarity threshold; the authors report up to 68% latency reduction with negligible accuracy loss. AxLLM is orthogonal to these reuse techniques and can be composed with them: when other methods eliminate work for some inputs, AxLLM applies computation reuse to the remaining multiplication operations on inputs that are not eliminated.

## III. AxLLM DESIGN

### a. Computation reuse opportunities

AxLLM accelerates the inference phase of transformers by applying computation reuse on matrix multiplication operations. Matrix multiplication is implemented in the *linear projection* and feedforward steps of each transformer layer. Linear projection transforms the input sequence into queries (Q matrix), keys (K matrix), and values (V matrix) by multiplying it to Wk, Wq, and Wv weight matrices, respectively. The feed-forward network consists of (often) two fully connected layers with a ReLU activation in between.

Fig. 1 outlines the computation load of all steps in each layer of a transformer. As the figure shows, the two operations we target dominate the layer computation, so reducing their complexity will translate to considerable reduction in the total transformer computation load.

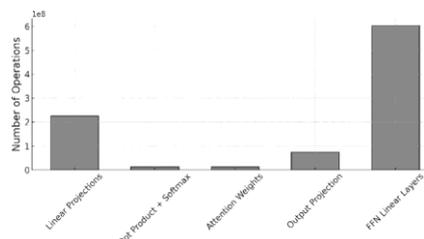

**Fig. 1. Contribution of each part to total computation in one layer of DistilBERT model**

### b. Computation reuse method

AxLLM exploits the value locality in weights, resulted from the limited precision (and hence limited unique values) introduced by quantization. The key idea is to identify and reuse specific partial sums of the matrix multiplication resulting from this value locality.

In an LLM, each layer receives a matrix input where each row represents the hidden state of a token (or the input embedding in the first layer). Focusing on the multiplication of a single row vector of the input matrix with the weight matrices, we adopt an input-stationary computation pattern, where the elements of the input vector (denoted x) are fetched and processed sequentially. In this scheme, the i-th element of x is multiplied with every element in the i-th row of the weight matrix W, contributing partial sums to the corresponding elements of the output vector. This flow is depicted in Fig. 2.

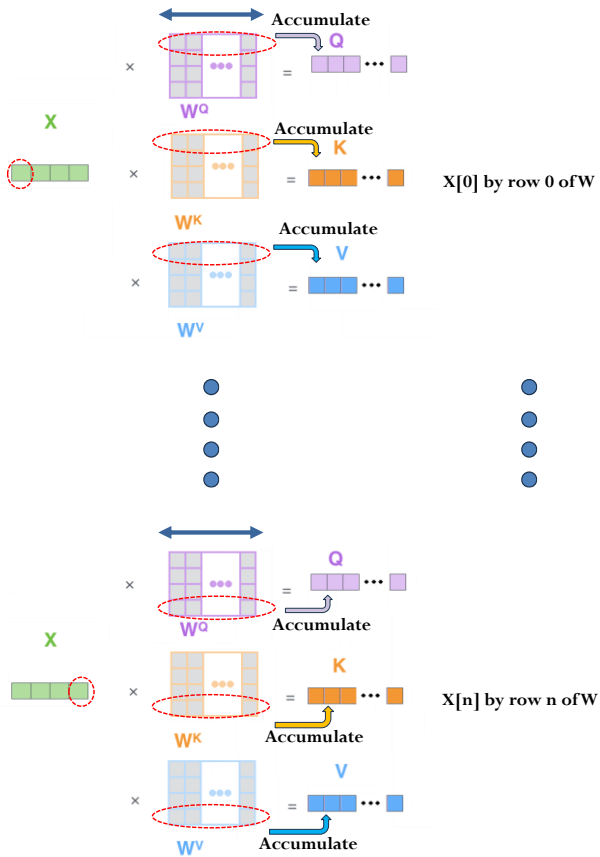

**Fig. 2. The input-stationary execution order of AxLLM**

For example, in Llama, a typical weight matrix in the self-attention layers is sized 4096×4096: with 8-bit quantization, the 4096 parameters in each row of W can take just 256 distinct values. Consequently, it is very likely that many repetitive values will appear in W rows. Computation reuse eliminates redundant multiplications arising from this repetition.

To eliminate redundant multiplications, we track all elements within a row that share identical values. In our value-centric, input-stationary approach, when the input cell x[i] is multiplied by the i-th row of W, the result of each unique weight value is cached upon its first occurrence. This cached result is then reused across all elements in the row that share the same value, eliminating the need for repeated computations.

To enable efficient computation reuse, the method incorporates a specialized Result Cache (RC). With q-bit quantization, the RC contains $2^q$ entries, each assigned to one of the unique values that the weights can take. With 8-bit quantization, which has been shown to strike an effective tradeoff between model complexity and accuracy loss [3], the RC buffer can be implemented with 256 entries.

Instead of absolute values, the weights in matrix W are now treated as pointers to the RC. The first occurrence of each unique value in a W row triggers a multiplication with the input, and the result is stored in the RC for subsequent reuse. The subsequent weights with the same value just pick up the result form RC and reuse.

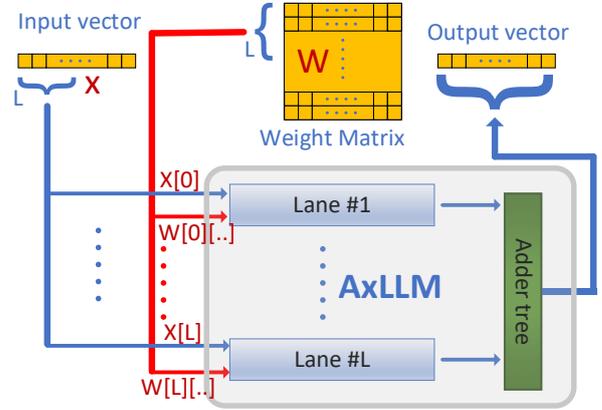

**Fig. 3. AxLLM architecture**

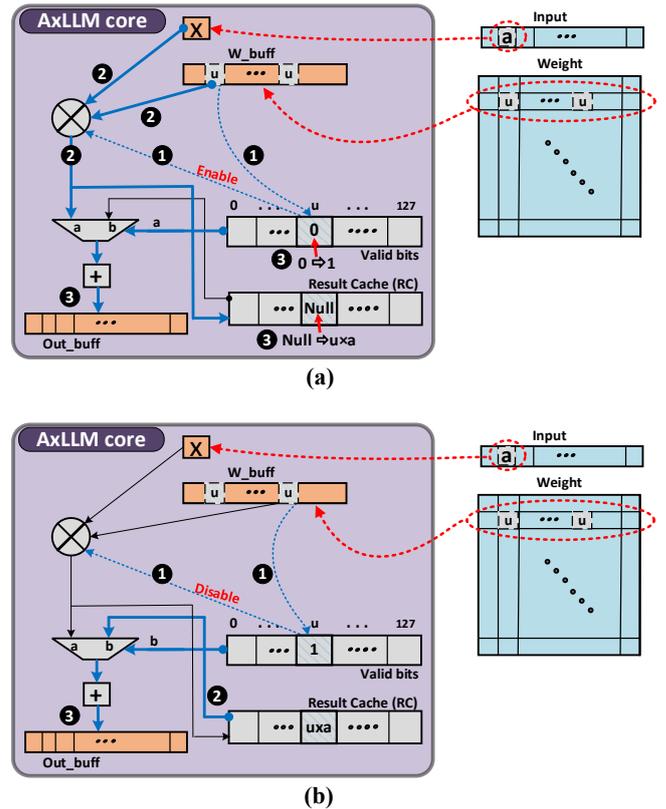

**Fig. 4. The architecture of a lane of AxLLM (a) compute and (b) reuse steps, at the first and subsequent occurrences of a unique weight value (u)**

*c. AxLLM architecture*

Fig. 3 outlines the AxLLM architecture. It composes of a set of L parallel lanes, where lane *i* multiplies the input cell x[i] to the i-th row of W. The vector of partial sums kept at each lane are accumulated using an adder tree and are stored to a global output buffer.

At each step, L elements of the input are fetched from input vector x and distributed to separate lanes. For each element x[i], sent to lane i, the corresponding elements of the i-th row of W are sent to the lane and streamed across it. Each lane implements one multiplier and two buffers, namely W_buff, and Out_buff, to keep the assigned row of the weights and the calculated partial sums, respectively. The input vector element assigned to the lane is stored in register X (Fig. 4).

The RC buffer, as the key enabler of the computation reuse, is the $2^Q$-entry result cache that keeps the multiplication result of the input cell kept in X and all the unique weight values of W_buff.

Fig. 4 shows the internal structure of each lane. Fig. 4.a. shows how the RC is filled and Fig 4.b shows how the RC contents are reused.

To manage this process within each lane, a controller performs the following: ❶ reads the weights of W_buff in order (let's say the value of the i-th weight is $u$) and checks if RC[$u$] contains a valid (already calculated) result. This is done by checking a valid flag of the RC[$u$].

If the flag is not set, the compute datapath will be invoked (Fig 4.a). In this path, ❷ the value of u×X (X keep s the input) is calculated. ❸ Then, the result is written to Out_buff[i], RC[u] is filled with u×X for future reuse, and the valid flag is set.

On the subsequent appearance of value u in W_buff, the reuse datapath will be taken (Fig 4.b). In this case, where the valid flag is set, ❷ RC[u] is read and ❸ written to Out_buff[i], bypassing the multiplier unit.

Once X produces all the partial sums, the results are sent out to the adder tree (Fig. 3) to accumulate with the output of other lanes. The RC is also cleared (by resetting the valid flags) and the algorithm continues with the next inputs.

**Feedforward layer support**. The feedforward layer, which accounts for the majority of computations in transformers (see Fig. 1), is highly amenable to computation reuse enabled by AxLLM. In the feedforward layer, the input vector is multiplied by the weight matrix of the neurons. Sharing the same structure of vector-by-matrix multiplication as the attention layer, the feedforward layer is seamlessly executed using the AxLLM datapath in our design.

**AxLLM support of LORA.** LORA-based fine-tuning, instead of updating all weights, introduces two low-rank trainable matrices, A and B, for each adaptable weight matrix W. In this way, the xW computation becomes xW + xAB, where x is the input vector, W is the original weight matrix, and A and B are low-rank matrices. While LoRA reduces memory usage by avoiding full weight storage for each task, it adds computational overhead due to the xAB calculation.

In addition to extracting computation reuse across the weight matrices, since both the main W and A are multiplied by x, AxLLM introduces a mechanism to share results between W and A.

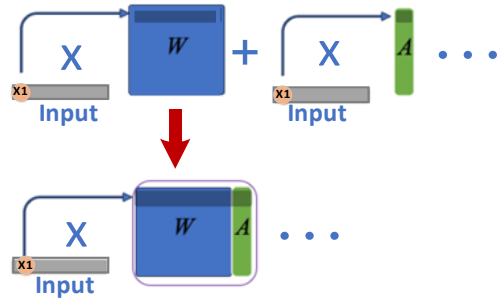

**Fig. 5. Combining W (weight matrix) and A (the first adaptor matrix) to extend computation reuse to A**

In this way, by treating the A matrix and W matrix as a single combined matrix—since they share the same number of rows but differ in column count—the results can be seamlessly reused across both matrices. Fig. 5 shows this scheme.

**Assumptions and Scope of Computation Reuse in AxLLM.** AxLLM works for the inference phase of LLMs. It does not change the model parameters, rather looks for equal weights to bypass their related computation, so does not need re-training.

It is important to note that in most LLMs, the input batch size and multi-head attention implementation alter the matrix dimensions. However, the core computational steps remain unchanged. To simplify the explanation, we describe the AxLLM architecture assuming a batch size and head size of 1. AxLLM can handle multiple attention heads and larger batch sizes by leveraging its existing cores, without modifying the core computation.

## IV. AxLLM DESIGN ISSUES

**Buffer size management**. AxLLM uses an input stationary order to enable computation reuse. However, this leads to incomplete computation of many output cells in the output buffers. To ensure scalability, we limit the input and output buffer size of each lane to 512 elements. This is the same size some outstanding related work that face the same issue, like the EIE accelerator [14], use. With this approach, rows of the weight matrix W are divided into parts of 256 to 512 elements each. The computation focuses on completing the first 512 columns of the output before moving to the next set of 512 columns. For example, in a 4096-column matrix (as in Llama), the execution is done in three steps (since 4096/512=8). In this way, the architecture completes the calculation of 512 cells of the output at each step, thus the number of incomplete results is restricted to 512 cells at any time. (Fig. 6)

This buffer size strikes a balance between area and data reuse. Larger buffers enable data reuse across the entire W row but increase chip area and complexity. Our experiments show that a buffer size of 512 offers a good compromise, providing sufficient reuse.

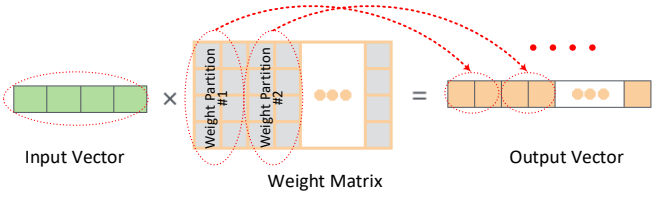

**Fig. 6. Restricting the weight matrix columns at each round (to 2 in this example) to reduce the buffering space required by incomplete output cells**

**AxLLM pipeline.** AxLLM already exploits parallelism at the input-cell level by assigning them to different lanes. Each lane consists of two pipeline paths—compute and reuse—that operate in parallel to improve throughput. Compute path processes the first occurrence of each unique weight value, and the reuse path accesses the result cache to retrieve the precomputed multiplication results for repeated weight values.

These two pipelines share the first and last stages: W_buff read and Out_buff write. Between these shared stages, the compute pipeline adds a multiplication step and the reuse pipeline adds an RC read stage.

According to the results, obtained by synthesizing the RTL description of the architecture in 15nm technology, we set the latency of the multiplier and buffer access stages to 3 and 1 cycles, respectively. The buffer size is set to 64, as will be explained shortly.

When a new weight element is fetched at cycle $t$, it passes through the multiplication path during cycles $t+1$ to $t+3$. Cycle $t+4$ is then available for writing back the multiplication result.

In parallel, subsequent weights are fetched each following cycle and are either directed to the reuse cache to retrieve cached results or queued for multiplication.

If a repeated weight is fetched at cycle $t+2$, its writeback will coincide with the writeback stage of the multiplication path from the weight fetched at cycle $t$, both targeting cycle $t+4$. This conflict is resolved by inserting a small queue for Out_buff, which will also be employed in the parallel-lane architecture described shortly.

By this design, computation reuse is done in parallel with multiplication.

To maintain pipeline efficiency, the system stalls only when the first appearance of a unique weight value V occurs at cycle t, and that same value V is required again in cycles $t+1$ to $t+3$. In such cases, the reuse path cannot proceed because the multiplication result for V is not yet available. While this hazard introduces potential stalls, our evaluation on our benchmarks (listed in Section 4) demonstrates that the likelihood of such occurrences is below 2%. As a result, the overall impact on performance is negligible, and more complex out-of-order weight processing mechanisms are not justified.

**Partitioning for Higher Throughput.** To increase throughput, the input, output, and reuse buffers are partitioned into S partitions, or slices. This allows one input to be read from each slice simultaneously, enabling P-way parallelism in every cycle (Fig. 7).

The controller supports P-way management, meaning it can fetch and process one input from each partition concurrently.

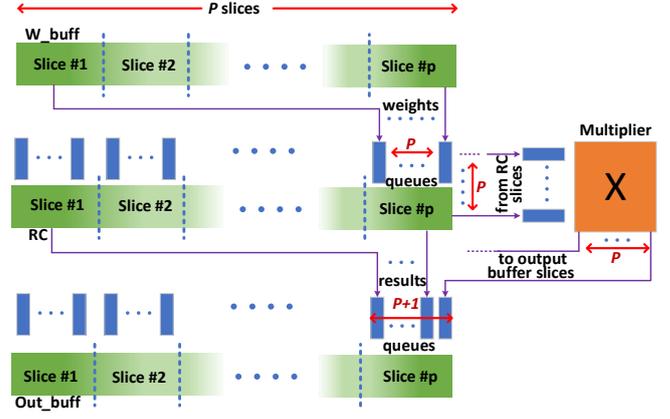

**Fig. 7. Parallel AxLLM architecture with n-way slicing of input buffer (W_buff), reuse cache (RC), and output buffer (Out_buff). The control unit along with each buffer slice is eliminated in the figure for the sake of simplicity**

**Collision Handling and Flow Control.** The highest throughput is achieved when the data fetched from all P partitions *(1) map* to different RC partitions, *and (2) each results* in a cache hit. In this case, the maximum possible number of results—P results—can be retrieved from the RC simultaneously. Since data from slice $i$ of W_buff always writes its result to the i-th slice of Out_buff, no output conflicts occur. As a result, the lane can sustain the maximum throughput of P partial sums in a single cycle.

Collisions can occur in the RC when multiple inputs refer to the same RC slice in the same cycle.

For example, input slices 1 and 2 may fetch weights with identical or close values at the same time, both requiring the partial result stored in RC slice 2.

To manage such conflicts, each RC slice and each output slice is preceded by a small queue (represented by blue blocks in fig. 7). With P input buffer slices, each RC slice has P queues (each of size S), and inputs are read in a round-robin fashion.

A credit-based back-pressure flow control mechanism is used between upstream and downstream buffers (e.g., between W_buff and the RC) to prevent writes to full queues.

In the worst-case scenario, where all weights fetched in the same cycle target the same RC slice, performance reverts to that of the non-parallel baseline AxLLM with unified buffers.

**Multiplier and Data Path Organization.** Each processing lane contains a single multiplier unit, as most computations are bypassed through the RC. The multiplier is fed by P queues, one from each RC slice. Each Out_buff slice has P+1 queues: P queues for outputs from RC slices and one queue for results from the multiplier.

The RC is implemented with a dual-port buffer, allowing one read and one write (by the multiplier) in the same cycle. The architecture ensures that reads and writes never target the same address within a cycle.

Fig. 7 shows the complete parallel AxLLM architecture, including buffer partitions, queues, and data flow lines between the RC, multiplier, and output buffers. The number of input and output slices must be the same.

While the RC could have a different number of slices, in our analysis we assume an equal count for simplicity and without loss of generality.

## V. EXPERIMENTAL RESULTS

**Simulation setup**. Table I summarizes the model/dataset pairs we use as benchmark. The models are taken from the HuggingFace pre-trained transformer library [15]. To evaluate the performance of our model on LORA adaptors, we also include two fine-tunned versions of BERT and DistilBERT as benchmarks.

The parameters of all models are quantized to 8-bit fixed-point signed numbers, at which all models keep the accuracy within 1% of the baseline. Since the weights are signed numbers, we maintain a 128-element reuse cache (instead of 256) and map each value and its negative to the same cell.

We built an in-house simulator that implements the AxLLM architecture, executes transformer-layer processing, and reports model reuse rate and runtime.

We also feed the activity factor of the models when running on the simulator to the VHDL model of AxLLM to estimate power usage.

TABLE I
DATASETS, TASKS, AND PRE-TRAINED MODELS USED

| Model | Dataset | Weight Matrix Size |
|---|---|---|
| DistilBERT | AG News | 768x768 |
| DistilBERT (fine-tunned) | Yelp Review Full | 768x768 |
| BERT Base Uncased | SQuAD | 768x768 |
| BERT Base Uncased (fine-tunned) | IMDb | 768x768 |
| Large BERT | IMDb | 1024x1024 |
| Llama 7B | IMDb | 4096x4096 |
| Lama 13B | IMDb | 5120x5120 |

**Computation reuse rate**. We first evaluate the computation reuse rate, i.e. the percentage of the multiplication results taken from the Reuse Cache across different layers of the models, and across the vectors (query, key, and value) in each layer. The weights are all quantized to 8-bit values. Fig. 8 shows that this rate is 87% at minimum.

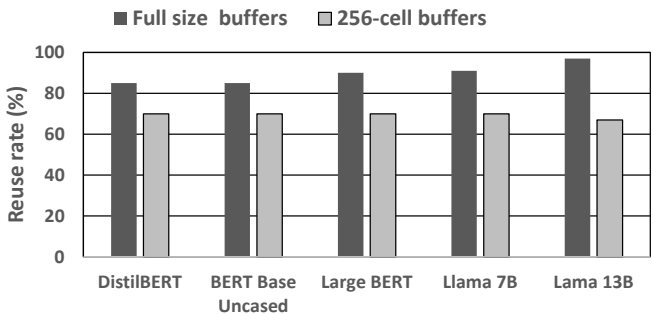

Fig 8. Reuse rate of the models

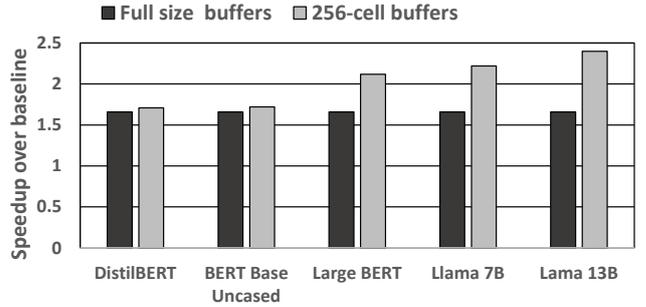

Fig 9. AxLLM speedup

The reuse rate grows with matrix size, as the set of unique values requiring multiplication is constant (128), whereas larger matrices introduce more additional multiplications that can be reused.

Fig. 8 also shows the reuse rate when the weight and output buffer sizes are limited to 256, reducing the overhead caused by incomplete cells. As the figure shows, all models achieve a similar reuse rate, averaging about 70%. We also observed that an average of 90% of the elements of each row of the adaptor matrix A repeats in the corresponding row in W, so the related multiplication can be reused to effectively eliminating part of the adaption computations.

Fig. 9 shows the execution time of the models when AxLLM is used. AxLLM is organized as a 64-lane architecture, each with 256-entry weight/output buffers. In each lane, the buffers are arranged as four 64-entry slices that are processed in parallel.

The results are normalized to a baseline architecture with the same size, where the AxLLM architecture with just multipliers (and not the reuse buffer) is used. For reference, the absolute value of the execution time of DistilBERT for AxLLM and the baseline are 85.11 and 159.34 million cycles, respectively.

As shown in the figure, AxLLM reduces execution time by an average of 1.7x. Since all models use the same buffer size, the reuse rate, and hence the speedup, converge to similar values.

Our results also shows that the execution speed related to the LORA adaptor matrices is increased by 1.82x BERT (fine-tunned for IMDb dataset) and 1.81x for DistilBERT (fine-tunned for *Yelp Review Full* dataset).

**Comparison with state-of-the-art**. We also compare our results with ShiftAddLLM [9], a state-of-the-art accelerator for transformers using one representative benchmark, DistilBERT. In ShiftAddLLM, with q-bit quantization, the weight matrix W is approximated by binary matrices $b_1, b_2, \ldots, b_q$, each containing +1 or −1, and scaling factors $\alpha_1, \alpha_2, \ldots, \alpha_q$, each rounded to a power of two to replace multiplication by shift. The multiplication of the activation vector x by W is approximated as $W.x = \sum_{i=1}^{q} a_i \cdot (b_i \cdot x)$. To further optimize, it precomputes the $256 (= 2^8)$ possible values for every 8-element sub-vector of the activations and stores them in a lookup table (LUT). Then, subsets of 8 elements from the binary matrices b are used as queries to the LUT, and the corresponding precomputed values are added together to produce the final output activations.

We implement ShiftAddLLM with 64 parallel shift-add units, matching the configuration of our 64-lane AxLLM. The results on 8-bit quantized DistilBERT as a representative benchmark reveal that AxLLM achieves a 29% speedup over ShiftAddLLM. ShiftAddLLM and AxLLM, when using the same quantization, require the same number of steps to complete vector-to-matrix multiplication. However, AxLLM achieves a performance gain primarily due to its design, which (1) enables parallel operations, and (2) eliminates the need for a setup phase to fill LUTs. Instead, its result cache is populated and utilized in parallel based on the occurrence of unique values.

**Power consumption**. To measure the power usage and area, we implement AxLMM in VHDL and synthesize the code using a publicly available 15nm library [16].
The results of running one layer of DistilBERT shows that with the AxLLM reuse, the average power consumption of the baseline is reduced from 0.94W to 0.67W. This is a result of replacing power-hungry multipliers with more power-efficient buffer reuse.

**Area**. Our synthesis results show that the AxLLM area is equivalent to 132k gates. The main components of the architecture are the input and output buffers, multipliers and accumulators, reuse cache, and controller, which account for 28%, 44%, 19%, and 9% of the total area, respectively. The area overhead is 23%, attributed to the reuse buffer and 4% of the controller area. This overhead is justified by a 1.7× increase in computation speed.

## VI. CONCLUSION

In this paper, we introduced AxLLM, an accelerator architecture that reduces the computational complexity of transformer-based LLMs and their related LoRA adaptors. AxLLM exploits computation reuse and value locality through low-bit quantization, eliminating redundant multiplications in weight and adaptor matrices. It employs a reuse cache to store multiplication results for unique weight values and reuses them when the same operands reappear. AxLLM adopts a parallel pipelined architecture to increase speed. Experiments show AxLLM reduces computations, achieving up to 28% power savings and 1.7x faster processing, demonstrating its potential as an efficient LLM solution.